\documentclass[aps,twocolumn,groupedaddress,showpacs]{revtex4}
\usepackage{graphicx}%
\usepackage{dcolumn}
\usepackage{amsmath}
\usepackage{latexsym}
\usepackage{epsfig}

\newtheorem{theorem}{Theorem}
\newtheorem{definition}{Definition}

\begin{document}

\title{Nonlocality, Asymmetry, and Distinguishing Bipartite States}

\author{Jonathan Walgate}
\email{jon.walgate@qubit.org}
\author{Lucien Hardy}
\email{lucien.hardy@qubit.org}
\affiliation{Centre for Quantum Computation, Clarendon Laboratory,
Parks Road, Oxford OX1 3PU, United Kingdom}

\date{February 5, 2002}

\begin{abstract}
Entanglement is an useful resource because some global operations cannot be locally implemented using classical communication. We prove a number of results about what is and is not locally possible. We focus on orthogonal states, which can always be globally distinguished. We establish the necessary and sufficient conditions for a general set of $2 \times 2$ quantum states to be locally distinguishable, and for a general set of $2 \times n$ quantum states to be distinguished given an initial measurement of the qubit. These results reveal a fundamental asymmetry to nonlocality, which is the origin of ``nonlocality without entanglement'', and we present a very simple proof of this phenomenon.
\end{abstract}


\maketitle

Many global operations cannot be performed using only local
operations and classical communication (LOCC). This one fact underpins
the use of entanglement as a resource across quantum information
theory, from teleportation \cite{teleport} to computation
\cite{comp}. Yet there is no clear delineation of what is and is
not locally possible. What evidence there is can appear
counter-intuitive, given the close link between entanglement and
nonlocal behaviour. Any two orthogonal entangled states can be distinguished
just as well using LOCC as they can globally \cite{Walgate} (see also \cite{Virmani,Chen}). But there exist larger sets of orthogonal separable states that
LOCC cannot reliably distinguish. Bennett et al presented a set of nine pure product states which, they proved, cannot be distinguished exactly with LOCC \cite{sausage}. (In fact they proved the stronger result that any approximate method to distinguish these states must introduce a finite error.)  Sets of states exhibiting such ``nonlocality without entanglement'' have been linked with unextendible product bases \cite{Bennett1, DiV}, in that the members of a UPB cannot be exactly LOCC distinguished. Yet this cannot be fundamental to the phenomenon, for there are LOCC indistinguishable sets that form complete orthogonal bases \cite{sausage}.

A set of states, shared between Alice and Bob, is exactly locally distinguishable if there is some sequence of local operations and classical communications that will determine with certainty which state they own. The Bell states present a simple example of an orthogonal set that is not locally distinguishable - a global measurement is needed to tell them apart
\cite{Ghosh1}.

In all LOCC protocols, one party must ``go first'' and perform the initial operation. We formalize this notion to obtain a powerful theorem establishing the necessary and sufficient conditions that a set of $2 \times n$ orthogonal quantum states are exactly distinguishable,
with the owner of the qubit going first (see definition 1, theorem 1 below). This result is of particular use in characterizing the distinguishabililty of $ 2 \times 2$ states. Recent
investigations by Ghosh et al, focusing on distillable entanglement, have revealed groups of orthogonal $2 \times 2$ states that are not LOCC distinguishable \cite{Ghosh2}. Using our result, we can now completely specify the distinguishable and undistinguishable $2 \times 2$ sets. 

Our theorem also provides insight into nonlocality without entanglement. Groisman and Vaidman \cite{Groisman} recently constructed a proof that Bennett et al's nine states cannot be exactly distinguished; they employed the idea that one party must go first, by considering results derived from a restriction to one-way communication. Theorem 1 allows us to develop a more transparent and natural proof of this important theorem, and go some way towards illuminating the origin of the phenomenon.

\begin{definition}
\textbf{Alice goes first} if Alice is the first person to perform
a nontrivial measurement upon the system.
\end{definition}
Note that this does not restrict two-way classical communication between Alice and Bob,
nor does it limit the number of measurements they may perform sequentially. Note also that in all LOCC protocols \emph{someone} goes first.

Consider Alice's first local operation.  Whatever she chooses to
do, whether she decides to involve an ancillary quantum system,
and whether she performs unitary operations as well as
measurements, Alice's actions will be described by a single
superoperator, $\$ $. The superoperator comprises a set $\{ M_{m}
\}$ of Krauss operators, one for every possible outcome, $m$. The
probability of a given state yielding a certain outcome is
\begin{displaymath}
p(m) = \langle \phi | M_{m}^{\dagger} M_{m} | \phi \rangle ,
\end{displaymath}
and the subsequent state of that system will be
\begin{displaymath}
\frac{M_{m} | \phi \rangle}{\langle \phi | M_{m}^{\dagger} M_{m} |
\phi \rangle}.
\end{displaymath}
The objects $M_{m}^{\dagger} M_{m}$ are the POVM elements
corresponding to each measurement outcome $m$. They sum to
identity. Being positive operators they are diagonalizable,
with real, nonnegative eigenvalues.  We will say that a
measurement is \emph{trivial} if all the POVM elements are
proportional to the identity operator since such a measurement
yields no information about the state. Any measurement not of this
type will be called \emph{nontrivial}.

\begin{theorem}
Alice and Bob share a $2 \times n$ dimensional quantum system:
Alice has a qubit, and Bob an $n$-dimensional system that may be
entangled with that qubit. If Alice goes first, a set of $ \ l$
orthogonal states $\{ | \psi_{i} \rangle \}$ is exactly locally
distinguishable if and only if there is a basis $\{ | 0 \rangle ,
| 1 \rangle \}_{A}$ such that in that basis:
\begin{gather}
\left| \psi_{i} \right\rangle = \left| 0\right\rangle _{A}\left| \eta_{0}^{i}\right\rangle_{B} + \left| 1\right\rangle _{A}\left| \eta _{1}^{i}\right\rangle _{B} \label{final} 
\end{gather}
where $\langle \eta_{0}^{i} | \eta_{0}^{j} \rangle = \langle \eta_{1}^{i} | \eta_{1}^{j} \rangle = 0 $ if $ i \neq j$.
\end{theorem}
Proof: The proof of sufficiency is simple. If there is a
basis such that the $l$ states can be written as above, the states
may be locally distinguished as follows. Alice measures in the $\{
| 0 \rangle , | 1 \rangle \}_{A}$ basis and communicates the
result to Bob. Bob then measures in the corresponding orthogonal
basis $ \{ | \eta_{0}^{i} \rangle \} $ or $ \{ | \eta_{1}^{i}
\rangle \}$, successfully distinguishing the states.

The proof of necessity is more complicated. Suppose that Alice
goes first.  The $l$ states must be reliably distinguished.
Therefore after each and every possible result of Alice's
measurement, all those states that have not been eliminated as
possibilities must remain orthogonal, and thus potentially
distinguishable. Therefore for all pairs of states $| \psi_{i}
\rangle $, $| \psi_{j} \rangle $, and for all measurement results
$m$, either that pair remains orthogonal post-measurement or else one of that pair of states has been eliminated.
\begin{align}
\textbf{Either} \ \ \ \ \langle \psi_{i} | M_{m}^{\dagger} M_{m} | \psi_{j} \rangle & = 0 \label{still orthogonal}, \\
\textbf{or} \ \ \ \ \langle \psi_{i} | M_{m}^{\dagger} M_{m} | \psi_{i} \rangle & = 0, \label{it's psi1} \\
\textbf{or} \ \ \ \ \langle \psi_{j} | M_{m}^{\dagger} M_{m} | \psi_{j} \rangle & = 0. \label{it's psi2}
\end{align}
Consider one POVM element that is not proportional to identity
(such an element must exist since Alice's measurement is
nontrivial), and take as our $\{ | 0 \rangle , | 1 \rangle \}_{A}$
basis the basis in which it is diagonal as follows:
\begin{equation} M_{m}^{\dagger} M_{m}=\left(
\begin{array}{cc}
\alpha & 0 \\ 0 & \beta
\end{array}
\right), \ \ \alpha > \beta\geq 0. \nonumber
\end{equation}
The states $|\psi_i\rangle$, expanded in the $\{ | 0 \rangle , | 1
\rangle \}_{A}$ basis at Alice's end can always be written in the form
of equation (\ref{final}). We must now prove the stated orthogonality conditions on the $|\eta\rangle$s.

For the moment consider only two states: $|\psi_i\rangle$ and
$|\psi_j\rangle$. If Alice eliminates neither of our pair from the running, those
states must remain orthogonal, in line with equation (\ref{still
orthogonal}). Since the original possible states are orthogonal as
well we require that:
\begin{equation} \langle \eta_{0}^{i} |
\eta_{0}^{j} \rangle + \langle \eta_{1}^{i} | \eta_{1}^{j} \rangle
= 0 , \ \ \ \alpha \langle \eta_{0}^{i} | \eta_{0}^{j}
\rangle + \beta \langle \eta_{1}^{i} | \eta_{1}^{j} \rangle = 0.
\nonumber
\end{equation}
These simultaneous equations combine thus:
\begin{equation}
(\alpha - \beta ) \langle \eta_{1}^{i} | \eta_{1}^{j} \rangle = 0 , \ \ \  \nonumber
(\beta - \alpha ) \langle \eta_{0}^{i} | \eta_{0}^{j} \rangle = 0. \nonumber
\end{equation}
Since $\alpha \neq \beta$
\begin{equation}
\langle \eta_{0}^{i} | \eta_{0}^{j} \rangle = \langle \eta_{1}^{i} | \eta_{1}^{j} \rangle = 0. \label{stuff4}
\end{equation}
Hence, in the case that neither state is eliminated, this pair of
states must be in the form given in the theorem.

Now consider the special case where Alice achieves a negative
identification by herself, in line with (\ref{it's psi1}) or
(\ref{it's psi2}). This tells us a great deal about that state.
Imagine she has eliminated $| \psi_{i} \rangle$. From
(\ref{final}) and (\ref{it's psi1}) we know that:
\begin{equation} \langle \eta_{0}^{i} | \eta_{0}^{i}
\rangle + \langle \eta_{1}^{i} | \eta_{1}^{i} \rangle = 1 ,
\ \ \ \alpha \langle \eta_{0}^{i} | \eta_{0}^{i} \rangle + \beta
\langle \eta_{1}^{i} | \eta_{1}^{i} \rangle = 0. \nonumber
\end{equation}
These simultaneous equations reveal that:
\begin{equation}
\alpha ( \langle \eta_{1}^{i} | \eta_{1}^{i} \rangle - 1 ) = \beta
\langle \eta_{1}^{i} | \eta_{1}^{i} \rangle. \label{stuff1}
\end{equation}
But $\alpha > \beta \geq 0$ and $ 0 \leq \langle \eta_{1}^{i} |
\eta_{1}^{i} \rangle \leq 1 $. This means that there is only one
possible solution to equation (\ref{stuff1}):
\begin{equation}
\beta =0  , \ \ \ \langle \eta_{1}^{i} | \eta_{1}^{i} \rangle
= 1. \nonumber
\end{equation}
This implies that $|\psi_{i}\rangle$ is the product state $
|1\rangle | \eta_{1}^{i} \rangle $. In this case, the other state
must take the form:
\begin{equation}
\left| \psi_{j} \right\rangle  = \left| 0\right\rangle_{A}\left|
\eta _{0}^{j} \right\rangle_{B}+ \left| 1\right\rangle_{A}\left|
\eta_{1}^{{i}{\perp}}\right\rangle_{B}. \label{stuff2}
\end{equation}
Again, we see that this particular pair of states have the form
given in the theorem.

Hence, in all cases, any pair of states must be in the form given
in the theorem. But the basis $\{ | 0 \rangle , | 1 \rangle
\}_{A}$ for which this is true depends only on the POVM element we
have been considering, and that element is independent of the
states themselves.  Therefore $\{ | 0 \rangle , | 1 \rangle
\}_{A}$ is a basis in which all the states are represented thus:
\begin{gather}
\left| \psi_{i} \right\rangle = \left| 0\right\rangle _{A}\left| \eta_{0}^{i}\right\rangle_{B} + \left| 1\right\rangle _{A}\left| \eta _{1}^{i}\right\rangle _{B} \nonumber \\
\textbf{where} \ \ \
\langle \eta_{0}^{i} | \eta_{0}^{j} \rangle = \langle \eta_{1}^{i} | \eta_{1}^{j} \rangle = 0 \nonumber \ \ \textbf{if} \ \  i \neq j. \nonumber
\end{gather}
This completes the proof. $\Box$
\bigskip

Theorem 1 depends upon the first measurement being made by the owner of the qubit. If we are dealing with $2 \times 2$ states, then the proof is applicable to both Alice and Bob going first. Thus any set of $2 \times 2$ states that can be locally distinguished must be expressible in form (\ref{final}). This allows us to derive the conditions for LOCC distinguishing all possible sets of orthogonal $2 \times 2$ states. Analysis has already shown that pairs of orthogonal states can always be LOCC distinguished:
\begin{theorem}
(Walgate et al.) Two orthogonal $2 \times 2$ states can always be exactly locally distinguished. \label{t2}
\end{theorem}
Proof: It was proved by Walgate et al \cite{Walgate} that Alice can always find a basis of form (\ref{final}) in which two states (of any dimension) can be distinguished. $\Box$
\begin{theorem}
Three orthogonal $2 \times 2$ states can be exactly locally distinguished if and only if at least two of those states are product states.
\end{theorem}
Proof: From theorem 2 it follows that any three states can be written thus:
\begin{align}
\left| \psi_{1} \right\rangle & = \left| 0\right\rangle _{A}\left| \eta
_{0}\right\rangle _{B}   + \left| 1\right\rangle _{A}\left|
\eta _{1}\right\rangle _{B}  , \nonumber \\
\left| \psi_{2} \right\rangle & = \left| 0\right\rangle _{A}\left| \eta
_{0}^{\perp}\right\rangle _{B}  + \left| 1\right\rangle _{A}\left|
\eta _{1}^{\perp}\right\rangle _{B}, \nonumber  \\\left| \psi_{3}
\right\rangle & = \left| 0\right\rangle _{A}\left| \nu
_{0}\right\rangle _{B}  +\left| 1\right\rangle _{A}\left|
\nu _{1}\right\rangle _{B}.  \nonumber
\end{align}
If this set is to be locally distinguishable with Alice going first, there must be some choice of $\{ |0\rangle , |1\rangle \}_{A}$ such that $\langle \eta_{0} | \nu_{0} \rangle = \langle \eta_{0}^{\perp} | \nu_{0} \rangle = 0$,
and $\langle \eta_{1} | \nu_{1} \rangle = \langle \eta_{1}^{\perp} | \nu_{1} \rangle = 0$. But there is no room in Bob's two-dimensional Hilbert space for three mutually orthogonal states. Therefore in each of these cases, one of the two (unnormalized) states forming the inner product must have zero magnitude. Since the states $|\psi_{i} \rangle$ must themselves be normalized, this means that two of them must be product states. This leaves us with the triplet:
\begin{eqnarray}
\left| \psi_{1} \right\rangle &=&\left| 0\right\rangle _{A}\left| \eta
_{0}\right\rangle _{B}+ \left| 1\right\rangle _{A}\left|
\eta _{1}\right\rangle _{B}  \nonumber \\
\left| \psi_{2} \right\rangle &=&\left| 0\right\rangle _{A}\left| \eta
_{0}^{\perp} \right\rangle _{B} \label{3} \\ \left| \psi_{3}
\right\rangle &=& \ \ \ \ \ \ \ \ \ \ \ \ \ \ \ \ \ \left| 1\right\rangle _{A}\left|
\eta _{1}^{\perp} \right\rangle _{B}.  \nonumber
\end{eqnarray}
Three orthogonal $2 \times 2$ states can be locally distinguished with Alice going first if and only if they take the form (\ref{3}). We can reconstruct this argument for Bob going first, with Bob's qubit providing the orthonormal basis. The form of states we obtain is a mirror image of (\ref{3}), with only the $ | \ \rangle_{A}$ and $ | \ \rangle_{B}$ indexes reversed. States of this form can still be locally distinguished, but now with Bob going first. It is easy to verify that these two arrangements encompass all sets of three orthogonal $2 \times 2$ states containing two product states. Therefore, three orthogonal $2 \times 2$ states can be locally distinguished if and only if at least two of those states are product states. $\Box$
\begin{theorem}
Four orthogonal $2 \times 2$ states can be exactly locally distinguished if and only if all of them are product states.
\end{theorem}
Proof: Given theorem 3, any three of a set of four distinguishable states must contain at least two product states. Thus two of $|\psi_{1}\rangle , |\psi_{2}\rangle , |\psi_{3}\rangle $, two of $|\psi_{1}\rangle  ,  |\psi_{2}\rangle  ,  |\psi_{4}\rangle $ and two out of $|\psi_{1}\rangle  ,  |\psi_{3}\rangle  ,  |\psi_{4}\rangle $ must be product states. It follows that at least three of the four states are product states - in general, for Alice going first, three such states can be written:
\begin{align}
|\psi_{1} \rangle &= |0\rangle_{A} |\phi \rangle_{B}, \nonumber \\
|\psi_{2} \rangle &= |1\rangle_{A} |\theta \rangle_{B}, \label{4}  \\
|\psi_{3} \rangle &= |0\rangle_{A} |\phi^{\perp}\rangle_{B}. \nonumber
\end{align}
There is only one state that is orthogonal to the above three, and that too is a product state:
\begin{equation}
|\psi_{4} \rangle = |1\rangle_{A} |\theta^{\perp}\rangle_{B}. \label{4.1}
\end{equation}
Four orthogonal $2 \times 2$ states can be locally distinguished with Alice going first if and only if all of them are product states of form (\ref{4}),(\ref{4.1}). Again, a complimentary argument with Bob going first provides another set of distinguishable product states, which together with set (\ref{4}),(\ref{4.1}) covers all possibilities. Therefore four orthogonal $2 \times 2$ states can be locally distinguished if and only if all of them are product states. $\Box$

A $2 \times 2$ system has a four-dimensional Hilbert space, and so cannot contain a set of more than four mutually orthogonal states. Thus this completes our analysis.

\bigskip

The sets of three and four LOCC distinguishable $2 \times 2$ states (\ref{3}),(\ref{4}) display a remarkable asymmetry: the states can be distinguished if one person goes first, but not the other way round.
\begin{definition}
A set of bipartite states is \textbf{asymmetrically} distinguishable if there is a specific party such that those states can only be exactly LOCC distinguished when that party goes first.
\end{definition}

Consider the triplet (\ref{3}). With Alice going first it is clear how to distinguish these states, but if Bob goes first this cannot be achieved because, as can be easily shown, there is no basis $\{ |0\rangle ,  |1\rangle \}_{B}$ in which the states take the form of theorem 1. This three-state asymmetry manifests if and only if one of the states is entangled. The corresponding four-state asymmetry, however, involves only separable states. 

The four states (\ref{4})+(\ref{4.1}) may be locally distinguished if Alice goes first, but not if Bob goes first so long as $| \langle \phi | \theta \rangle | \neq 1$. Bob can do nothing reliable until he receives some information from Alice. Conversely, Alice can only reliably discover which state she possesses by allowing Bob to discover, and hoping that he shares his knowledge. An example of this phenomenon was discussed by Groisman and Vaidman \cite{Groisman}. They imposed stronger constraints, limiting Alice and Bob to one way communication in the direction $B \rightarrow A$, and showed that the states given by $|\phi\rangle_{B} = |0\rangle_{B} \ , \ |\theta\rangle_{B} = |0+1\rangle_{B}$ cannot be distinguished in that circumstance. This is a four-state example of the asymmetry we have outlined, which arises not from a one-way communication restriction, nor indeed any practical limits on Alice and Bob's LOCC protocol. Rather, this asymmetry appears in the set of states itself: Alice and Bob will know without consultation who must make the first move.

This asymmetry emerges from the most basic level. The two states $|0\rangle_{A} |\phi\rangle_{B}$ and $ |1\rangle_{A} |\theta\rangle_{B}$ are of course orthogonal, but whilst Alice's intervention is both necessary and sufficient to distinguish them, Bob's is not. The point is the orthogonality of any pair of product states must be \emph{locally} manifested. One might naively expect that the addition of the second pair of orthogonal states, ``completing'' the $2 \times 2$ Hilbert space, would provide a balance, and reintroduce symmetry. This is not the case.

There is one and only one ``symmetric'' set of four orthogonal $2 \times 2$ states, in the sense that there is only one set that is reliably discriminated no matter who measures first:
\begin{equation}
|0\rangle_{A} |0\rangle_{B}  , \ \
|1\rangle_{A} |0\rangle_{B}  , \ \
|0\rangle_{A} |1\rangle_{B}  , \ \
|1\rangle_{A} |1\rangle_{B}. \nonumber
\end{equation}
An interesting property of these states is that they can encode a single bit such that neither Alice nor Bob can access it without help from the other:
\begin{eqnarray}
\textbf{Let} \ \ |0\rangle_{A} |0\rangle_{B} \ \ \textbf{and} \ \ |1\rangle_{A} |1\rangle_{B} \ \ \textbf{encode ``0''}. \nonumber \\
\textbf{Let} \ \ |0\rangle_{A} |1\rangle_{B} \ \ \textbf{and} \ \ |1\rangle_{A} |0\rangle_{B} \ \ \textbf{encode ``1''}. \nonumber
\end{eqnarray}
Both Alice and Bob have the power to reveal the bit to their partner, but neither can gain any access to it directly.
\bigskip

Nonlocality without entanglement occurs when a set of product states can not be distinguished with either Alice or Bob going first. Bennett et al's paper considered a set of nine such states, which were symmetric under the exchange of Alice and Bob's systems. But this symmetry is not fundamental to the nonlocality. In its simplest form, we can think of nonlocality without entanglement manifesting asymmetrically for only for one party. Groisman and Vaidman used this insight when they created a proof of Bennett's result built from their observations on one-way indistinguishability \cite{Groisman}. What is really at issue is not the kind of LOCC protocols employed by Alice and Bob, nor the content of their communications, but the asymmetric properties of subsets of the states themselves. Framed this way, the ``full-blown'' phenomenon has a very simple proof.

\begin{figure}[h]
\centerline{\includegraphics{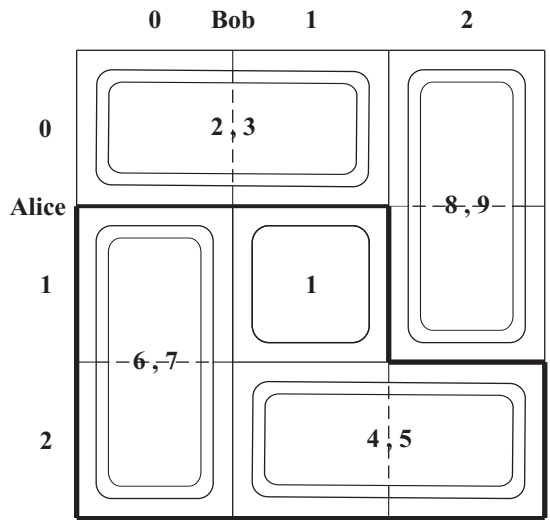}}
\caption{Bennett et al's depiction of the states (\ref{sausages}) as a set of dominoes.}
\end{figure}

\begin{theorem}
(Bennett et al) The nine $3 \times 3$ states depicted in figure 1 and specified below cannot be exactly distinguished using only local operations and classical communication.
\begin{align}
|\psi_{1} \rangle &= |1\rangle_{A} | 1 \rangle_{B} \nonumber \\
|\psi_{2,3} \rangle &= |0\rangle_{A} | 0\pm1 \rangle_{B} \nonumber \\
|\psi_{4,5} \rangle &= |2\rangle_{A} | 1\pm2 \rangle_{B} \label{sausages} \\
|\psi_{6,7} \rangle &= |1\pm2\rangle_{A} | 0 \rangle_{B} \nonumber \\
|\psi_{8,9} \rangle &= |0\pm1\rangle_{A} | 2 \rangle_{B} \nonumber
\end{align}
\end{theorem}
Proof: We will prove that the states cannot be distinguished if Alice goes first. If so then by their symmetry the states cannot be distinguished with Bob going first either.

Alice performs a general measurement, represented by a set of $3 \times 3$ POVM elements $M_{m}^{\dagger}M_{m}$, which we will write in the $\{ |0\rangle, |1\rangle, |2\rangle \}_{A}$ basis:
\begin{equation}
M_{m}^{\dagger} M_{m}=\left(
\begin{array}{ccc}
m_{00} & m_{01} & m_{02} \\
m_{10} & m_{11} & m_{12} \\
m_{20} & m_{21} & m_{22}
\end{array}
\right) \nonumber
\end{equation}
The effect of this positive operator upon states 1,4,5,6 and 7 (highlighted in bold in the diagram) is entirely specified by those elements drawn from the $\{ |1\rangle, |2\rangle \}_{A}$ subspace: $m_{11}, m_{12}, m_{21}$ and $m_{22}$. This select set of states is of dimension $2 \times 3$, yet there is palpably no basis in which Alice can express them in the form of theorem 1. These states are thus indistinguishable with Alice going first, and Alice cannot perform a nontrivial measurement upon the $\{ |1\rangle, |2\rangle \}_{A}$ subspace. Thus the corresponding sub-matrix must be proportional to the identity, and hence $m_{11} = m_{22}$ and $m_{12} = m_{21} = 0$.

Exactly the same argument can be made for the states 1,2,3,8 and 9 and the $\{ |0\rangle, |1\rangle \}_{A}$ subspace. Therefore $m_{00} = m_{11}$ and $m_{01} = m_{10} = 0$. Since $M_{m}^{\dagger}M$ is Hermitian, $m_{20} = m_{02}^{\ast}$. Alice's POVM element must look like this:
\begin{equation}
M_{m}^{\dagger} M_{m}=\left(
\begin{array}{ccc}
\alpha & 0 & m_{02} \\
0 & \alpha & 0 \\
m_{02}^{\ast} & 0 & \alpha
\end{array}
\right) \label{simple}
\end{equation}
Now consider the $\{ |0\rangle, |2\rangle \}_{A}$ subspace, and the states 2 and 4. Alice's measurement must either leave them orthogonal or distinguish them outright. In the former case, we demand that $\langle \psi_{4} | M_{m}^{\dagger} M_{m} | \psi_{2} \rangle = 0$. Simple algebra shows that, given (\ref{simple}), $\langle \psi_{4} | M_{m}^{\dagger} M_{m} | \psi_{2} \rangle = \frac{1}{2} m_{02}^{\ast}$. Thus $m_{02} = 0$ and $M_{m}^{\dagger} M_{m}$ is proportional to the identity.

If Alice distinguishes the states outright then for one of states 2 and 4, $\langle \psi_{i} | M_{m}^{\dagger} M_{m} | \psi_{i} \rangle = 0$. But given (\ref{simple}), $\langle \psi_{i} | M_{m}^{\dagger} M_{m} | \psi_{i} \rangle = \alpha$. Thus $\alpha = 0$ and, since POVM elements must be positive, $M_{m}^{\dagger} M_{m}$ is the null matrix.

The above argument applies to all possible measurement outcomes, and thus all of Alice's POVM elements must be proportional to the identity if she and Bob are to distinguish the states. By definition, Alice cannot go first. By the symmetry of states (\ref{sausages}), neither can Bob. Therefore the states (\ref{sausages}) cannot be distinguished using only local operations and classical communication. This completes the proof. $\Box$

\bigskip

We have shown that sets of orthogonal $2 \times n$ states can be distinguished only if they can written in a particular form, and we have seen how this result dictates the distinguishability of the $2 \times 2$ states. ``Nonlocality without entanglement'' can be constructed from the asymmetries that arise in sets of such states.

\bigskip

We would like to thank the UK \mbox{EPSRC} and the Royal Society for funding this research.

\end{document}